# An Ear Canal Deformation Based User Authentication Using Ear Wearable Devices

ZI WANG, Florida State University, USA
SHENG TAN, Trinity University, USA
LINGHAN ZHANG, Florida State University, USA
YILI REN, Florida State University, USA
ZHI WANG, Florida State University, USA
JIE YANG, Florida State University, USA

Biometric-based authentication is gaining increasing attention for wearables and mobile applications. Meanwhile, the growing adoption of sensors in wearables also provides opportunities to capture novel wearable biometrics. In this work, we propose EarDynamic, an ear canal deformation based user authentication using in-ear wearables. EarDynamic provides continuous and passive user authentication and is transparent to users. It leverages ear canal deformation that combines the unique static geometry and dynamic motions of the ear canal when the user is speaking for authentication. It utilizes an acoustic sensing approach to capture the ear canal deformation with the built-in microphone and speaker of the in-ear wearable. Specifically, it first emits well-designed inaudible beep signals and records the reflected signals from the ear canal. It then analyzes the reflected signals and extracts fine-grained acoustic features that correspond to the ear canal deformation for user authentication. Our extensive experimental evaluation shows that EarDynamic can achieve a recall of 97.38% and an F1 score of 96.84%. Results also show that our system works well under different noisy environments with various daily activities.

## 1 INTRODUCTION

Recently, the new-generation in-ear wearables have integrated various sensors (e.g., microphones, vibration and motion sensors) to provide better user experience and to support a wide range of emerging applications. For example, the Apple AirPods Pro equips multiple outward-facing and inward-facing microphones to improve

Authors' addresses: Zi Wang, Florida State University, 1017 Academic Way, Tallahassee, FL, 32306, USA, ziwang@cs.fsu.edu; Sheng Tan, Trinity University, One Trinity Place, San Antonio, TX, 78212, USA, stan@trinity.edu; Linghan Zhang, Florida State University, 1017 Academic Way, Tallahassee, FL, 32306, USA, lzhang@cs.fsu.edu; Yili Ren, Florida State University, 1017 Academic Way, Tallahassee, FL, 32306, USA, ren@cs.fsu.edu; Zhi Wang, Florida State University, 1017 Academic Way, Tallahassee, FL, 32306, USA, zwang@cs.fsu.edu; Jie Yang, Florida State University, 1017 Academic Way, Tallahassee, FL, 32306, USA, jie.yang@cs.fsu.edu.

the ability of the active noise cancellation [6]. The Huawei Freebuds embeds bone vibration sensors to enhance the quality of voice recording [1], whereas the Jabra Sport earbud integrates motion sensors to support fitness monitoring [2]. These embedded sensors in the wearables can not only support various mobile applications but also provide opportunities to sense and capture new types of biometrics. For example, it is possible to utilize the embedded sensors in the in-ear wearable to capture a user's ear canal structure, which is unique to each individual for user authentication. Moreover, many emerging applications that enabled by the in-ear wearable, such as the voice assistant for smart home and IoT devices, require secure user authentication to protect sensitive and private information. Traditional voice-based user authentication is convenient but has been proven vulnerable to the voice spoofing attack [21, 22, 49]. Reusing in-ear wearable to capture the ear canal structure for user authentication thus provides a novel and promising approach to enhance system security.

Comparing to traditional biometrics, the ear canal based authentication has several advantages. First, it relies on the unique geometry of the ear canal, which is hidden within the human skull. It is thus more resilient to the spoofing attack than the traditional biometrics, such as fingerprint, voice, or face. For example, the victim's voice can be recorded by an attacker within the vicinity using a smartphone [22], whereas the facial authentication can be spoofed using the photos or videos of the victim [30]. Moreover, the ear canal based approach can provide continuous authentication, whereas the fingerprint and face biometrics mainly focus on one-time authentication. In addition, the ear canal based authentication is transparent to the users as it doesn't require any user cooperation. Traditional biometrics, however, require explicit user operation, such as pressing the fingertip on the fingerprint reader or posing the face to the camera [8, 28]. The unique advantages of the ear canal based authentication could potentially benefit many emerging applications, such as Virtual Reality (VR) and Audio Augmented Reality (AAR). For example, AAR utilizes earbuds to provide the user with acoustic information that corresponds with the physical reality and is widely used in museum or event tour. Wearing earbuds is a prerequisite for a better immersive experience but not an additional burden for the user. In VR, due to the view of the user is completely blocked from the environment, simply utilizing traditional authentication methods are very challenging. For example, existing authentication methods become less secure and more vulnerable to various attacks under the VR setup. Similarly, using hand gestures to input PIN in a virtual keyboard is vulnerable to video-recording attack while voices sample could be easily recorded at a close distance for the mimic attack without user knowledge. Ear canal based authentication is thus both convenient and more secure for those applications.

In this paper, we introduce EarDynamic, a continuous user authentication system that leverages the ear canal deformation sensed by the in-ear wearables. Specifically, the ear canal deformation reflects the ear canal dynamic motion caused by jaw joint or articulation activities, for example, when the user is speaking. Thus, the ear canal deformation not only contains the static geometry of ear canal that represents the physiological characteristic of the user but also includes the geometry changes that characterizes the behavioral properties of the user while speaking. Recent prior work shows that the static geometry of the ear canal is unique for every individual [15]. We find that the ear canal deformation due to articulation activities includes more dynamic information and could provide better and more secure user authentication while the user is speaking.

In particular, our system utilizes an acoustic sensing approach to capture the ear canal deformation with the embedded microphone and speaker of the in-ear wearable. It first emits inaudible beep signals from the in-ear wearable to probe the user's ear canal. It then records back the reflected inaudible signals together with the user's audible sounds. The reflected inaudible signals thus contain the information of the ear canal dynamic motions, i.e., the changes of the geometry of ear canal, due to the movements of the jaw joint and other articulators. More specifically, human speech relies on the motions of multiple articulators (e.g., jaw, tongue, mouth) to pronounce various phonemes. When the jaw moves, the temporomandibular joint (TMJ) also moves which causes either expansion or compression of the ear canal wall. Such a phenomenon caused by TMJ movements is known as the Ear Canal Dynamic Motion (ECDM) [12]. Such motions would result in various effects over different people [32]. For example, during the speaking process, each person's ear canal will either be expanded or compressed at various degrees and speeds. We found through the study that the ear canal deformation is consistent for the same individual while varies depending on the individual anatomy and behavior. Thus, by utilizing the ear canal deformation extracted from the reflected acoustic signals, our system can distinguish different users.

To better leverage the ear canal deformation and improve the usability of our system, we categorize various dynamic motions into different groups based on the phoneme pronunciations. In particular, although we cannot directly measure the canal dynamic motions, we can infer such motions based on the articulatory movements. However, measuring the articulatory movements requires specialized sensors attached to the articulators, which are impractical. We solve such a challenge by looking at the phonemes in the user's speech. Specifically, each phoneme pronunciation corresponds to unique and consistent articulatory movements. The canal dynamic motions thus could be identified by recognizing each phoneme and the corresponding articulatory movements. Consequently, the ear canal dynamic motions for the phonemes that are invoked by similar jaw and tongue movements will share high similarity and could be categorized into the same group. Such a categorization can also reduce the calculation complexity and shorten the authentication time.

To perform user authentication, our system extracts fine-grained acoustic features that correspond to the ear canal deformation and compares these features against the user enrolled profile. To evaluate EarDynamic, we conduct experiments with 24 participants in various noisy environments (i.e., home, office, grocery store, vehicle, and parks). We also evaluate our system under different daily activities during the user authentication process (i.e., maintain different postures, perform different gestures). The results show EarDynamic achieves high accuracy and maintain comparable performance in different noisy environments and under various daily activities. The contributions of our work are summarized as follows:

- We show that the dynamic deformation information of the ear canal is unique for each individual and can be utilized for user authentication. We further categorize ear canal dynamic motions into various groups based on the phoneme pronunciation to facilitate user authentication.

- We proposed EarDynamic, a continuous and user transparent authentication system utilizing in-ear wearable devices. It leverages ear canal deformation that combines the unique static geometry and dynamic motion of the ear canal when the user is speaking for authentication. A prototype of EarDynamic is built with off-the-shelf accessories by embedding an in-ward facing microphone inside an earbud.

- We conduct extensive experiments to evaluate the performance of the proposed EarDynamic. Experimental results show that our system achieves a recall of 97.38% and an F1 score of 96.84%. Results also show that EarDynamic works well under different noisy environments with various daily activates.

## 2 RELATED WORKS

### 2.1 Biometric Based Authentication

The commonly used biometrics, such as fingerprint, palmprints, face, and iris, rely on physiological characteristics. The fingerprint-based authentication has been incorporated in many forensic, governmental, and civilian applications [13, 27]. Recently, it has been quickly adopted on mobile and IoT devices. However, the fingerprint is vulnerable to malicious attacks, such as fingerprint obfuscation and impersonation [28, 29]. Moreover, it could be blotted or mimicked by removing or replacing part of the skin on the fingertip. For example, researchers found that artificial fingerprints could be obtained from latent fingerprints using silicon and moldable plastic [3] to deceive the recognition systems [40]. Similarly, palmprints can be utilized for authentication [19, 20] but also suffer from spoofing attacks. The face is another widely used physiological characteristics based authentication. Early commercial applications utilizing face recognition dated back to the early 1990s by extracting features from still images. Facial authentication now has been supported on mobiles. For example, smartphones manufactured by Apple and Samsung all include face authentication features [8]. However, a recent study suggests that face recognition has become more suspectable to presentation attacks due to the usage of deep neural networks [30]. Meanwhile, it is vulnerable to face morphing attacks [14]. Another problem with facial authentication is it will not work when the face is covered (e.g wearing a mask under COVID-19). Under such scenarios, the iris [26] could be an alternative. For example, Raja *et al.* [33] proposed a system utilizing deep sparse filtering to achieve iris recognition on the smartphone. However, iris-based authentication requires specialized hardware and it is vulnerable to high-quality image spoofing attacks.

The other type of authentication modalities focuses on behavioral characteristics, such as voice, signature, gait, and keystroke. Among them, voice has become one of the most popular biometric on smartphones and IoT devices due to the popularity of the voice assistance systems. However, recent studies found that the voice biometric is vulnerable to spoofing attacks [21, 22, 42], such as replay attacks [18] and speech-synthesize attacks [49]. Signature has been heavily adopted in the legal and banking system. It utilizes the distinct features of an individual's natural behavior of signing. However, the signature dynamic could be forged by professionals and also subjects to inconsistent even within the same individual. Gait is a behavioral biometric that captures the user's anatomy and the motion features of the gait [47]. The existing study also shows that daily activities could be one of the behavioral characteristics. For example, Shen *et al.* [41] utilize multiple motion sensors on mobile devices to perform continuous smartphone authentication. However, gait and activity-based authentication lack reliability and are very vulnerable to mimic attacks [11,45]. Keystroke is another behavioral characteristic that leverages the user's typing behavior for authentication [41]. However, it can be difficult to adopt due to large intra-variance attributed to a large number of factors, such as user behavior and type of keyboard.

## 2.2 Ear Canal Sensing

Researchers have been exploring the dedicated devices to sense the ear canal to performing various applications [46,50]. For example, researchers have been using piezoelectric ear canal sensors to infer the heart rate [31], leveraging the pulse oximetry sensor to sense the ear canal for monitoring pulmonary [44], and utilizing in-ear EEG sensor inside the ear canal to sense the brain waves [16]. Moreover, researchers also proposed methods to measure the canal deformation information. For example, Carioli *et al.* and Bedri *et al.* leverage piezoelectric sensors and infrared proximity sensors to estimate the ear canal deformation caused by the jaw movements, respectively [9, 10]. However, these methods utilizing specialized devices involve significant cost and deployment overhead. In this work, we aim to leverage the built-in sensors of the wearable to sense the ear canal deformation without requiring dedicated devices.

Another important application of sensing the ear canal is for user authentication. For example, Arakawa et al. [7] proposed a user authentication system utilizing microphone-integrated earphones to capture the static ear canal geometry. It extracts the Mel-frequency cepstral coefficients (MFCCs) features of the reflected acoustic signals from the ear canal to distinguish different users. Similarly, EarEcho [15] captures the acoustic characteristics of the static geometry of the ear canal for user authentication. These studies show strong evidence that the static geometry of the ear canal is unique to each individual. However, they only extract the static information of the ear canal but ignore the dynamic deformation associated with various activities, such as when a user is speaking or moving his/her head. In particular, the geometric of the human ear canal stay consistent when there is no head motions or mouth movements. However, when the user is turning his/her head (e.g, nodding, shaking) or moving the jaws (e.g., eating, speaking), these motions will lead to dynamic changes in the ear canal geometry, which is referred to as ear canal deformation. The recent work recognizes facial expressions by analyzing the reflected ultrasounds from the user's ear canal [5]. It achieves a F-score of 62.5% for 21 facial expressions and 90% for 6 facial expressions. However, the proposed system mainly focuses on recognizing facial expressions for the hands-free input method. Consequently, these approaches either only work well under static scenarios where no user activities are involved or do not focus on authentication purposes.

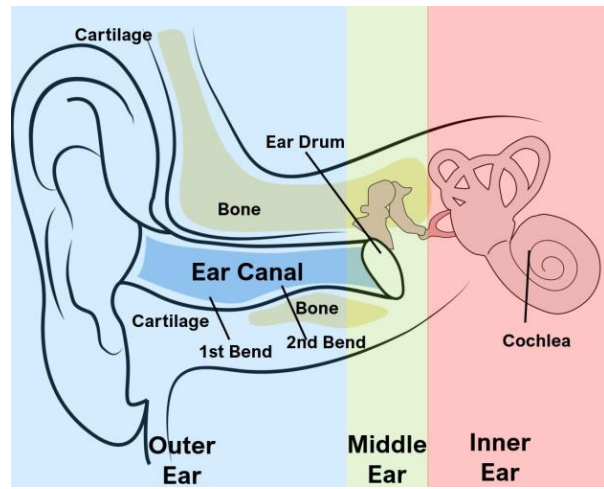

Fig. 1. The auditory system of the human ear

## 3 PRELIMINARIES

### 3.1 System and Attack Model

Our system utilizes dynamic deformation of the ear canal captured by the in-ear wearable for user authentication. It requires the in-ear wearable equipped with one microphone and one speaker. The authentication could be continuous, which means the microphone keeps sending inaudible probe signals for continuous authentication. Or the authentication could be triggered on-demand based on the requirements of the mobile applications. The authentication process is also transparent to the users as it doesn't require any user cooperation. During authentication, the user is free to speak, conduct activities, or remain silent. If the user is speaking or conducting activities, the dynamic ear canal deformation will be extracted for authentication.

In this work, we consider two types of attack models: mimic attacks and advanced attacks. For the mimic attacks, an adversary attempts to compromise the system by spoofing the ear canal deformation of the legit user. In particular, an adversary wears the victim's in-ear device and tries to issue the voice commands for user authentication. Moreover, the attacker might mimic the jaw or head motions of the legitimate user to bypass the user authentication. The second type of attack is an advanced attack. Although the ear canal is hidden within the human skull, it is still possible such information could be leaked, for example, through the 3D ear canal scanning when the victim needs to produce a hearing-aid or cure ear disease. For this type of attack, we assume an adversary acquires the user's ear canal geometry information and could rebuild it precisely, such as using 3D-printing technology. Such attacks may easily bypass the system that only utilizes the static geometry of the ear canal for user authentication. However, it is extremely hard, if not possible, for an adversary to reproduce the dynamic motions of the ear canal to bypass the ear canal deformation based authentication.

## 3.2 Geometry of Ear Canal

The ear is the auditory system of humans, which consists of three parts, outer ear, middle ear, and inner ear, as shown in Fig. 1. The ear canal together with pinna makes up the main components of the outer ear. The ear canal has a tube-like structure and ends at the outer surface of the eardrum. Its primary function is transmitting and coupling sounds. There exists a wide variety of geometry of the human ear canal across the population due to its geometric complexity. These varieties can be attributed by three aspects: ***interspace*** of the ear canal, the ***curvature and cross-section*** of the canal, and the ***composition*** of the ear canal wall.

Firstly, the ***interspace*** of the ear canal varies from one individual to another [34, 36]. As shown in Fig. 1, the ear canal of the adults is an "S" shape like a tube. It usually has two bends with different angles that divide the whole canal into three sections. For different individuals, the variations lie in different aspects such as length, size, and shape of the ear canal. For the length, a prior study [34] suggests that the length of the ear canal could vary from 14.20 mm to 29.36 mm, while the standard deviation of the ear canal length is approximately 2 mm among males and is 5% longer for females on average. Moreover, the volume of the ear canal's interspace ranges approximately from 1000 to 1400 $mm^3$ [36], and other recent studies involve more subjects reports that the smallest interspace could be as low as 372 $mm^3$ with a mean volume of 843 $mm^3$ [34].

Moreover, both the ***curvature and cross-section*** of the canal are also unique for different people. The curvature is usually measured along the center axis of the ear canal. The ear canal of certain individuals could be straight but are bent for others. In one study involving 185 adults suggests that for about 30% of the subject being studied, the entire eardrum could be seen from the point of view near the pinna. On the other hand, about 9% of testing subjects' eardrum is invisible from the viewpoint at the same position, which indicates these ear canals are relatively narrow and curved [25]. Furthermore, the cross-sectional area changes over the entire course of the ear canal due to its complex structure. For example, in the middle portion of the canals, the cross-sectional areas can range between 25 and 70 $mm^2$ for different subjects.

Lastly, the ***composition*** of the ear canal wall is skin, cartilaginous, and bone where the proportion of cartilaginous and bone are different for each person. As shown in Fig. 1, part of the ear canal wall is mainly the cartilaginous and the remaining parts are the bones. The ear canal wall consists of two portions that separated with an osteocartilaginous junction. The first one is the outer cartilaginous portion, where the ear canal wall is constituted with skin and cartilaginous. This portion usually comprises 1/3 to 1/2 of the total length of the ear canal and varies from one individual to another. The other portion is the inner osseous portion, where the ear canal wall mainly consists of skin and bone. It is worth noticing that the tympanic membrane, the junction of the outer ear and the middle ear, may also be considered as part of the ear canal wall. The yielding ear canal wall is more likely to absorb energy than those consists of skin and bone. Compare to the floor of the canal, the eardrum terminates the ear canal in a range of angles from 45° to 60° [36].

## 3.3 Dynamic Deformation of Ear Canal

Because human speech relies on the motions of articulators including jaw, tongue, and lips, it also causes the dynamic deformation (i.e., expansion or compression) of the ear canal, known as the Ear Canal Dynamic Motion (ECDM). Such motions mainly consist of ***mouth motion*** and ***head motions***.

Specifically, ***mouth motion*** includes both ***jaw and tongue movement***, where the geometry of the ear canal changes when people are speaking or chewing. As shown in Fig. 2, this is due to the temporomandibular joints (TMJ), the two joints connect the jaw with the skull are located near the ear canal walls. When people are speaking or chewing, the jaw movement drives the temporomandibular joints to move, thus further shifting the ear canal wall and changing the geometry of the ear canal. Furthermore, the ***head motions*** will also impact the shape of the ear canal. The sternocleidomastoid muscle, as shown in Fig. 3, is one of the largest and strongest neck muscles attached to the skull near the ear canal. When the head is turning from side to side, the sternocleidomastoid muscle will compress or expand the ear canal.

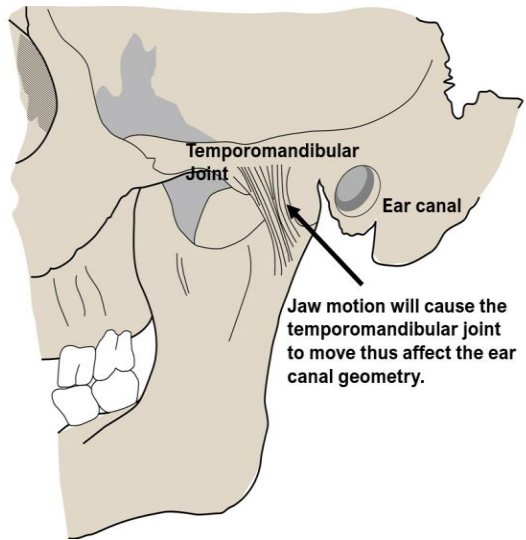
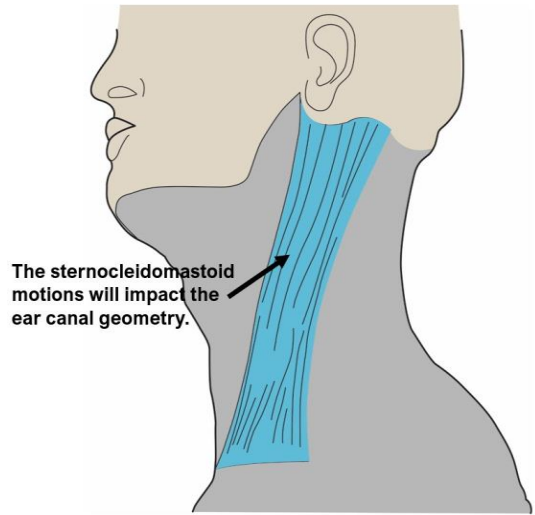

Fig. 2. Temporomandibular joints of the human skull

Fig. 3. The sternocleidomastoid muscle

One important observation is that the dynamic deformation of the ear canal also has diversity. Research [32] shows that with the same motion of open mouth (e.g., drop the jaw), the volume changes vary from person to person. The ear canal volume will be compressed as high as 10 $mm^3$ for about 25% of the testing subjects and expanded as high as 25 $mm^3$ for 67% of the subjects. Meanwhile, the volume of the remaining 8% of the subjects' ear canal keeps the same even when the joints move. The ear canal diameters also change differently for different individuals. An experiment involving 1488 ears suggests that with a similar joints motion, about 20% of the subject's ear canal diameters decrease while the remaining subjects increase up to 2.5 $mm$. Some studies [17] also show that the ear canal moves anteriorly, posteriorly, or with a combination of both for different people. In our work, we confirm that the dynamics of the ear canal have individual differences. Thus, the ear canal deformation caused by articulation activities could be leveraged for user authentication.

### 3.4 Ear Canal Deformation Categorization

To better leverage the ear canal deformation and improve the usability of our system, we categorize various ear canal dynamic motions into different groups. As the ear canal deformation is invoked by the movements of the articulators (e.g., jaw, tongue, and mouth) when the user is speaking, we thus can categorize deformation by measuring various articulator gestures. But such measurements usually require multiple specialized devices attached to the articulators, which are impractical and cumbersome. To solve this technique challenge, we rely on the phoneme pronunciation. In particular, each phoneme pronunciation corresponds to unique and consistent articulatory gestures. By grouping the phonemes with similar articulatory gestures, we are able to group the corresponding ear canal deformation. Thus, we first categorize the phonemes with a similar scale of jaw and tongue movements into the same groups. Then, the corresponding ear canal deformations belong to the same group as well due to the deformation are caused by similar jaw and tongue movements. We then refine such grouping by experimenting on different subjects and re-categorize the phonemes/deformations with high similarity.

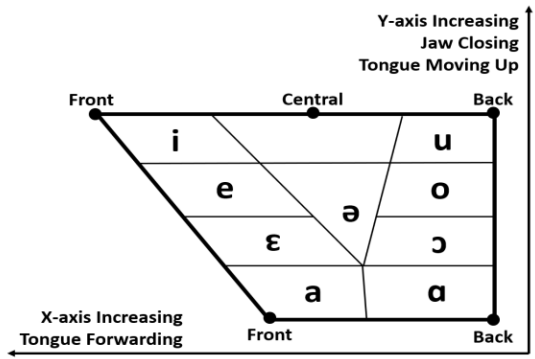 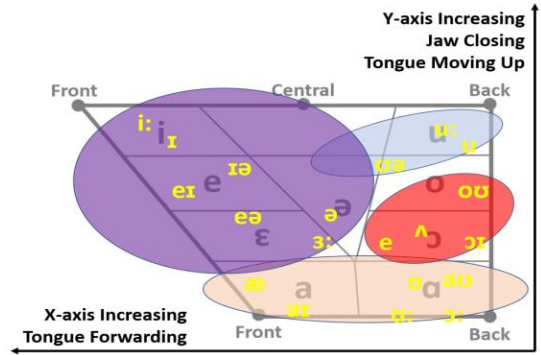

Fig. 4. Articulatory Phonetics: Jaw and tongue position     Fig. 5. Vowels categories based on the jaw and tongue position

Specifically, phonemes are the smallest distinctive unit sound of a language and can be divided into two major categories: vowels and consonants. Based on the position of the jaw and tongue when pronouncing them, we can categorize phonemes into different groups. Fig.4 illustrates the articulatory phonetics correspond to the various position of tongue and jaw. The y-axis shows the height information of the speaker's tongue. It indicates the vertical placement of the tongue at various degrees, which also associated with the jaw openness. The x-axis represents the depth information, which implies the forward or backward level of the tongue within the mouth. And the vowels are categorized based on the Articulatory Phonetics, as shown in Fig.5. For instance, the phoneme [i] on the left-up corner of the chart is a vowel, where the jaw is closed and the tongue is at a high and forward position closed to the palate and tooth during pronunciation. On the other hand, the phoneme [a] on the left lower corner is another vowel, where the jaw is open and the tongue is at a low and back position when pronouncing it. Similarly, we could categorize the consonants based on the position of the jaw and the tongue, as shown in Fig.6. Specifically, the x-axis shows the vertical placement of the tongue and the y-axis represents the depth information.

### 3.5 Ear Canal Deformation Sensing

Measuring the geometry of the ear canal directly is challenging due to its structural complexity. Specifically, instead of being a straight tube, the ear canal is an "S" shaped cavity and its cross-section keeps changing along the entire canal. Moreover, due to the complex articulatory gestures, the dynamic deformation caused by these gestures is also difficult to capture directly. For example, the jaw motion can result in various deformation on the ear canal in different directions. The measuring process becomes even more difficult when combining the static geometry with the dynamic deformation of the ear canal.

In our work, we use an acoustic sensing approach to capture the ear canal deformation indirectly. It is done by emitting inaudible acoustic signals to probe the ear canal and analyzing the signal reflections that affected by various geometric characteristics of the ear canal (e.g., the interspace, the curvature, the diameter of the ear canal, and the canal wall composition). We build a prototype using off-the-shelf ear-bud equipped with an inward-facing microphone that enables both transmitting probe signals and recording the signal reflections from the ear canal. Moreover, we leverage the channel response to measure the ear canal dynamic deformation and its geometry information. Specifically, the channel response of the ear canal is the ratio of the reflected signal to the incident probe signals. The channel response depicts how the ear canal deformation (i.e., wireless channel) reflects the original probe signals. By analyzing the channel response, we thus could distinguish different users based on their unique ear canal deformation.

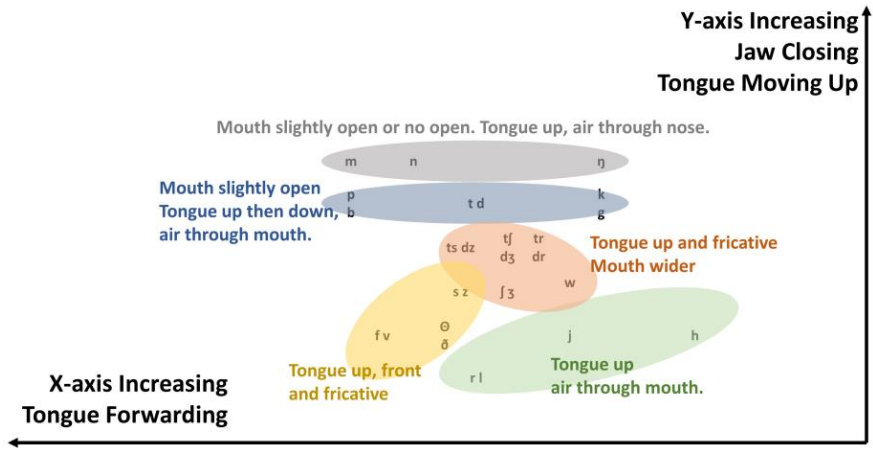

Fig. 6. Consonants categories based on the jaw and tongue position

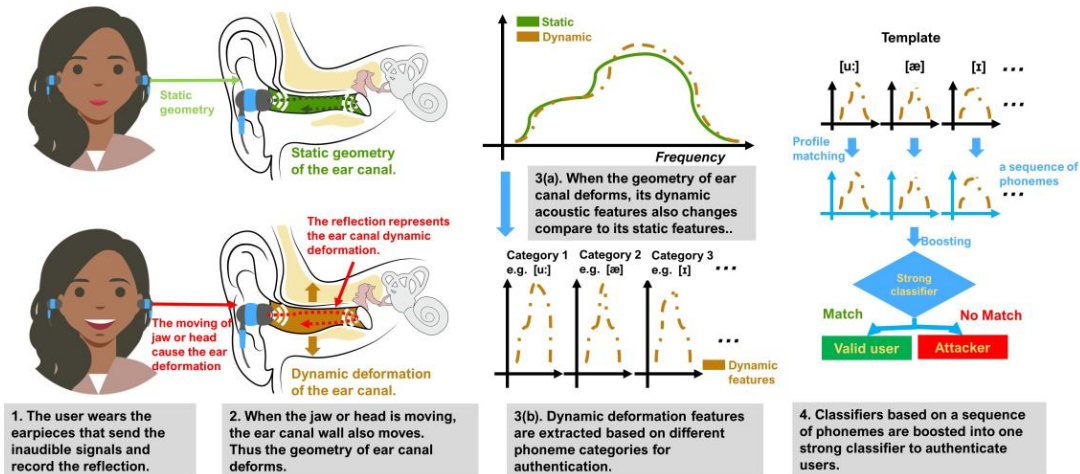

Fig. 7. The core idea of the EarDynamic

## 4 SYSTEM DESIGN

In this section, we introduce our system design and algorithms.

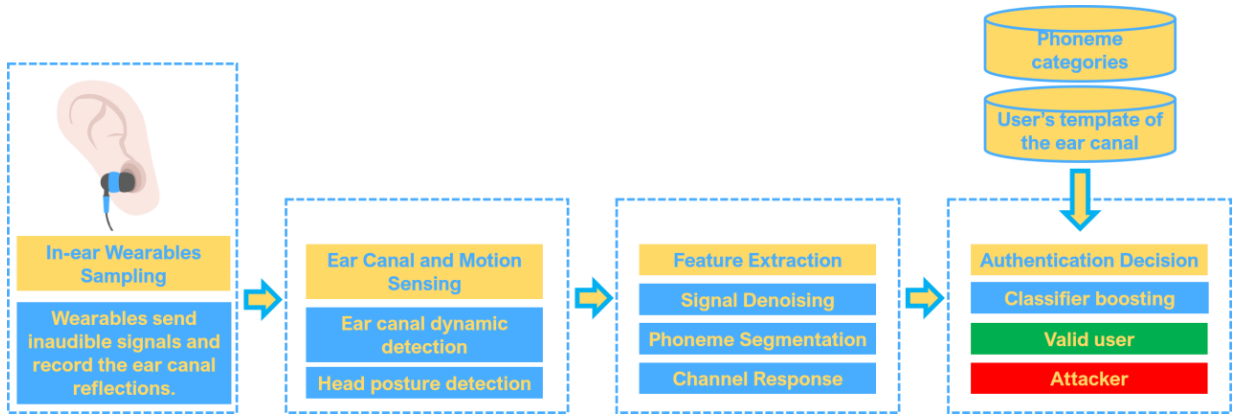

Fig. 8. System flow of EarDynamic.

## 4.1 Approach Overview

The key idea underlying our user authentication system is to leverage the advanced acoustic sensing capabilities of the in-ear wearable device to sense the dynamic deformation of the user's ear canal. As illustrated in Fig. 7, when the user is wearing EarDynamic, the earbud will emit an inaudible chirp signal to probe the ear canal. Then the signal reflected from the ear canal will be captured by the inward-facing microphone that can be further utilized to extract the dynamic deformation of the ear canal.

Our system has the ability to work under both static and dynamic scenarios of the ear canal. When there is no head movements or articulatory motions detected, the user is under a static scenario, where the ear canal geometry remains the same throughout the authentication process. Thus, the captured signal reflections represent the physiological characteristics of the user's ear canal. Then the extracted features that correspond to static geometry of the ear canal are utilized to compare against the user enrolled profiles to determine if it is the legit user.

Different from the static scenario, dynamic deformation represents the combination of both the physiological and behavioral characteristics of the ear canal. It can be extracted under the dynamic scenarios, where the user is speaking or moving the head. To better leverage the ear canal deformation, we categorize various deformation motions into different groups based on phoneme pronunciation, such that each group shares similar jaw and tongue movements. Such an approach has the benefit of improving system usability by simplifying the profile enrollment process. For example, our system only requires the user to speak a few sentences that involve multidimensional motions of the jaw and tongue to generate an individual profile for later authentication. To identify the head movement, our system relies on the embedded motion sensor of the wearable. In particular, we consider five head postures that lead to ear canal deformation including turning right, left, down, up, and forward.

Our system can perform authentication when users wearing in-ear devices with their natural habits. By utilizing embedded inward-facing microphone and motion sensor, which have been increasingly adopted on the wearable devices, EarDynamic is highly practical and compatible. Compare with the traditional biometric authentication modalities, such as fingerprint and face, our system can achieve continuous authentication and is transparent to the user without requiring any user cooperation.

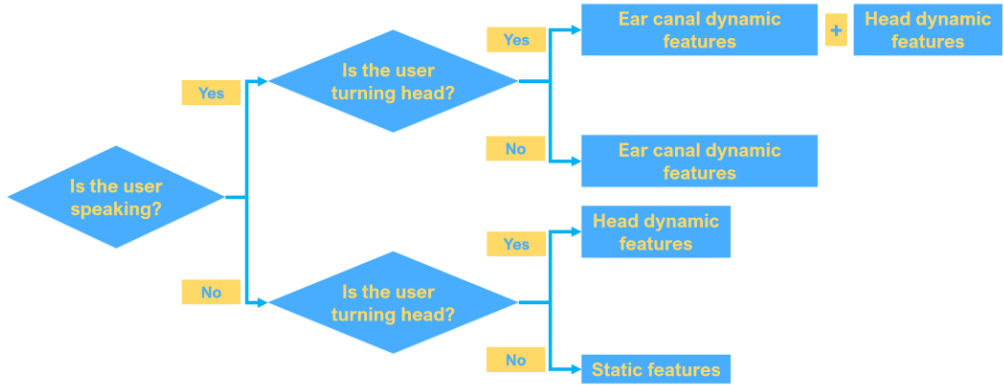

Fig. 9. Decision tree of choosing deformation template.

## 4.2 System Flow

Our system consists of four major components: in-ear wearable sampling, ear canal, and motion sensing, dynamic feature extraction, and user authentication, shown in Fig. 8. The authentication process could be triggered on-demand or continuously depending on the applications. Once triggered, our system will first send the probe signals and record the signal reflections. Meanwhile, the audible signal that contains phoneme information of the user's speech will also be recorded. Furthermore, the motion sensor will capture the head movements. Then, the system will move on to the feature extraction component.

For feature extraction, we need to process the captured signals that consist of the reflected inaudible signals and the audible signals. We first apply a high pass filter to separate the inaudible and audible signals. The inaudible signals contain the acoustic properties of the ear canal, whereas the audible signals include the speech content of the user. Next, our system segments the separated audible signals into a sequence of phonemes and maps them to the corresponding inaudible components for capturing the ear canal deformation. The phoneme segmentation is done by utilizing the Munich Automatic Segmentation system [39] on both the audible signal components and inaudible signal components. For each segment, we need to extract appropriate features based on different scenarios, as shown in Fig. 9. In particular, if the user is speaking while moving his/her head, both the ear canal dynamic and head dynamic feature will be extracted, respectively. On the other hand, if there is no head motion or mouth movement, the static geometry of the ear canal will be extracted.

Lastly, our system will authenticate the user based on the extracted information from previous steps. We utilize a sequence of phoneme-based classifiers that can be combined as one stronger classifier to improve classification accuracy. If a positive decision is given, then the user is considered as legit. Otherwise, our system will deem the current user as an unauthorized user.

## 4.3 Ear Canal Sensing

Once the system is triggered, the speaker of the in-ear wearable sends out acoustic signals to probe the ear canal. The probe signals are designed as a chirp signal with frequency ranges from 16kHz to 23kHz [50]. The reason for such a design is twofold: first, the frequency range from 16kHz to 23kHz is inaudible to most human ears, which makes the authentication process transparent to the user; second, the chosen frequency range is sensitive to the subtle motions, which can improve our system's ability to capture the ear canal deformation. During the process of ear canal sensing, the inward-facing microphone will keep recording the reflected signals bounced from the ear canal. These reflections can be analyzed to further extract the acoustic properties of the ear canal.

Table 1. Deformation Categories Based on Phoneme

| Deformation (Articulator) Category | Phonemes |
| --- | --- |
| Tongue Forward and Jaw Open Slightly | [i:], [I], [I@], [eI], [@], [e@], [3:] |
| Tongue Lower and Jaw Open Widely | [æ], [ai], [6], [A], [O:], [au] |
| Tongue Back and Raise and Jaw Open Slightly | [U], [u:], [U@] |
| Tongue Back and Jaw Open Moderately | [oU], [OI], [e], [2] |
| Tongue Raised and Fricative and Jaw Open Widely | [tS], [tr], [ts], [dZ], [dr], [dz] |
| Tongue Raised and Jaw Open Slightly | [f], [s], [S], [h], [v], [z], [Z], [r] |
| Tongue Fricative and Jaw Open Slightly | [T], [D], [l] |

### 4.4 Signal Processing

**Ear Canal Deformation Categorization.** The underlying principle of ear canal deformation categorization is that similar articulatory gestures, i.e., jaw and tongue motions, will have a similar impact on the geometry of the ear canal. In particular, each phoneme is produced by a sequence of coordinated movements of several articulators. In this work, we mainly focus on two articulators (i.e., jaw and tongue) that contribute the most to the ear canal deformation. For example, the phoneme sound of [O:] and [A] both have a lower and backward position with an open jaw. Thus, these two phonemes result in a similar impact on the ear canal deformation and are categorized into the same group. We also eliminate several phonemes due to the fact that they have minimal usage of articulators, which leads to almost no impact on the ear canal deformation. For instance, when the user pronounced the phoneme [p], no ear canal deformation was detected. The categorization results of commonly used vowels and consonants are summarized in Table 1.

As each phoneme contains unique formats (e.g., frequencies), we thus could segment and identify each phoneme by analyzing the spectrogram of the audible signal. In particular, we first leverage the automatic speech recognition protocol to identify each word in the sample speech [35]. Then, we utilize MAUS as the primary way of phoneme segmentation and labeling [24]. It is done by transferring the samples into expected pronunciation and searching for the highest probability in a Hidden Markov Model [23]. The segmented and labeled phonemes will be categorized according to Table 1 for further analysis.

**Feature Extraction.** The next step is to extract features from the categorized signals segments. The captured signal reflection from the ear canal contains the acoustic characteristics information of the user's ear canal. However, due to the dynamic nature of the ear canal geometry during the speaking process, the channel response of the received signals are also time-varying [43].

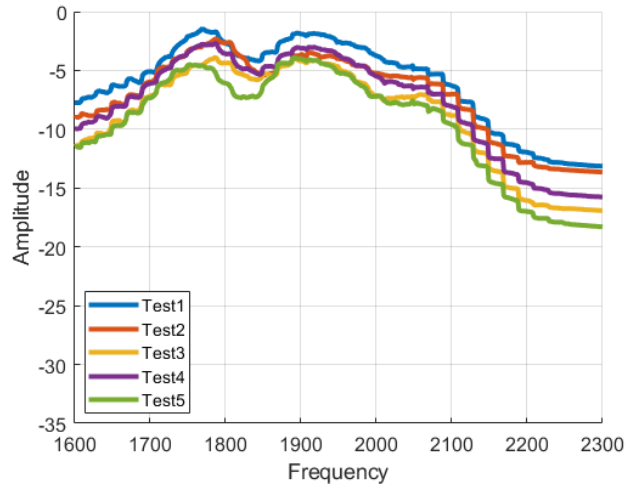

Fig. 10. Channel response of the same ear canal deformation category for same participant.

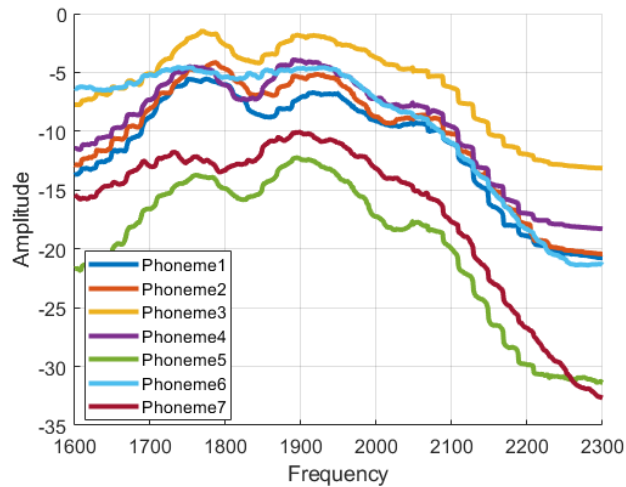

Fig. 11. Channel response of different ear canal deformation categories for same participant.

By capturing the channel response under different segments of phonemes, we could extract the acoustics characteristics that represent both the static geometry and dynamic motions of the ear canal at a specific time point. As shown in Fig. 10, the channel responses of different phonemes within the same category collected from the same participant share a high similarity. Furthermore, the channel responses of the phonemes across different categories have lower similarity even when they are collected from the same individual, as illustrated in Fig. 11. Lastly, we observe that the channel responses of the phonemes from the same phoneme category collected from different users are unique, as depicted in Fig. 12.

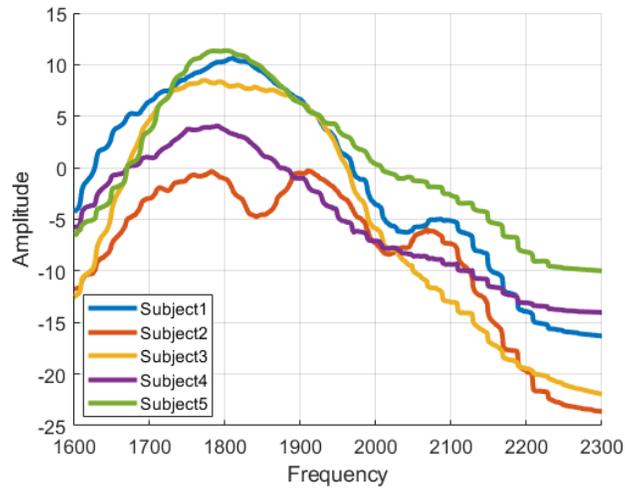

Fig. 12. Channel response of the same ear canal deformation category for different participants.

## 4.5 Classifier Boosting

After obtaining the feature extracted from received reflected signals, our system proceeds to the authentication process. In this work, such a process can be viewed as finding an optimal solution for a classification problem to distinguish between legit users and attackers [51]. To achieve better performance, we adopt adaptive boosting which is an ensemble learning algorithm [37,38]. Such an algorithm is commonly used for classification or regression to further improve the distinguishing ability. Specifically, the authentication problem can be formulated as one boosted classifier.

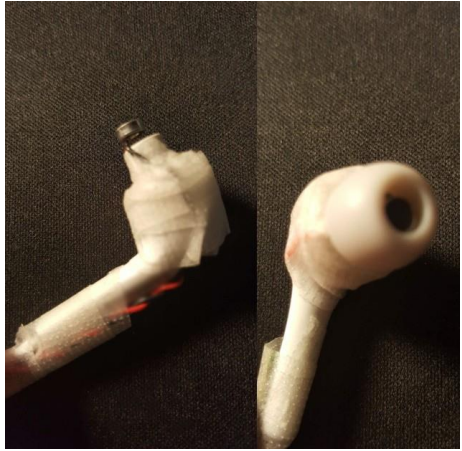

Fig. 13. The prototype built with off-the-shelf devices.

## 5 PERFORMANCE EVALUATION

### 5.1 Experimental Setup

**Environments and Hardware.** The authentication process could be happening in various environment under everyday use scenarios. For example, the user might command the voice assistant through the headset in the office environment. Additionally, the user could make payments at groceries using electronic payment or send a message through voice command in the vehicles. Thus, to evaluate our system's performance in real-world environments, we choose various locations including home, office, grocery store, vehicle, and parks to conduct experiments. Moreover, to better simulate the everyday usage of EarDynamic, we ask the participants to wear our system in their natural habits as to how they would wear in-ear devices on a daily basis. The participants are allowed to maintain various posture (sitting, standing, walking) or perform different gestures (e.g., waving arm and hands, moving head) during the experiments.

There are several in-ear earbuds that equipped with inward-facing microphones on the market (e.g., Apple Airpods Pro [6], Amazon EchoBuds [4]). However, those devices are less desirable due to firmware restriction and we can not access the raw data for feature extraction. In this work, we built our prototype system utilizing only off-the-shelf hardware to demonstrate its practicability and compatibility. We select a regular in-ear earbud on the market cost less than 7 dollars with a 12mm speaker, 3.5mm audio jack, and a microphone chip with a sensitivity of -28±3 dB. The total cost of this prototype is very low which is more affordable to a wider range of customers compared to the abovementioned earbuds. As shown in Fig. 13, the microphone is attached to the earbud where located in front of the speaker, and kept in the center of the cross-section area. Such design can mitigate the impact of wearing the earbuds in different angles while gives enough surrounding space for the probe signals to propagate through. We use two different smartphones including Samsung Galaxy Note 5 and Galaxy S8+ that connect to the prototype to control probe signal emitting and reflected signal recording. To detect the head posture, existing wearable devices use motion sensors such as accelerometer or gyroscope. For our system, we employ MPU6050, a six-axis motion sensor to detect the user's head position. The motion sensor is attached to the earbud and connected to a Raspberry Pi for data transmission and power.

**Data Collection.** We recruit 24 participants for the experiments including 12 females and 12 males with an age range from 20 to 40. The participants are informed about the goal of our experiments and asked to talk in their natural way of speaking. For the enrollment, each participant is asked to sit in a classroom environment and wear the prototype at his/her habitual position. Then, they are required to repeat five passphrases three times while our system emitting inaudible signals. Such passphrases are designed to include different dynamic deformation motions from all the categories. Then, the features are extracted from the captured signal reflections along with the audible signals. The extracted features are used to establish each user's template. After enrollment, each participant is asked to speak 10 sentences with length varies from 2 to 20 words. The sentences include some commonly used voice commands like "Hey Google", "Alexa, play some music" as well as other short daily conversation pieces. For each sentence, each participant is asked to repeat at least 10 times for our experiment. In total 2880 sentences from different users are collected and used for overall evaluation. Among all those sentences, 1080, 700, 300, 300, 300, and 200 sentences are collected at home, in the classroom, in office, in the vehicle, at the grocery store, and park respectively.

**Metrics.** To better evaluate the authentication performance of our system, we introduced four different metrics: accuracy, recall, precision, and F1 score.

We also leverage the receiver operating characteristic (ROC) which represents the relationship between the True Accept Rate (i.e., the probability of identified valid user) and the False Accept Rate (i.e., the probability of incorrectly accept the attacker) when the threshold is varying.

### 5.2 Overall Performance

We first evaluate our system's overall performance against the mimic attack. To launch a mimic attack, the adversary will wear the in-ear device and issue the same voice command by mimicking the victim's way of speaking. For such an attack, the adversary tries to spoof the system by performing similar articulator gestures with respect to the victim. Table 2 summarizes the average and median of accuracy, recall, precision, and F1 score overall. We can observe that EarDynamic can achieve overall accuracy of 93.04%, recall of 97.38%, the precision of 95.02%, and an F1 score of 96.84% across different environments and participants. Furthermore, the median accuracy, recall, precision, and F1 score are 93.97%, 98.78%, 95.40%, and 96.85%, respectively. Fig. 14 shows the ROC curves of our system with and without phoneme boosting. By utilizing the phoneme boosting technique, our system could provide over 95% True Accept Rate with around 5% False Accept Rate. The above results show that EarDynamic is highly effective in distinguishing legit users and attackers under mimic attack, especially after applying phoneme classifier boosting techniques.

When applying classifier boosting techniques, our system requires to utilize multiple phonemes to boost classification accuracy. Next, we study our system's performance by using the various number of phonemes as classifiers, the results are shown in Fig. 15. Overall, we observe that our system could achieve considerable accuracy by only using a few phonemes. Specifically, with just one phoneme being used, our system achieves over 89% accuracy, 94% recall, precision, and F1 score. Furthermore, the accuracy, recall, precision, and F1 score is improved to around 95%, 95%, 97%, and 98% respectively with only five phonemes. This result shows that EarDynamic could provide high authentication accuracy using only a few words for classifier boosting.

Table 2. Authentication accuracy

|  | Mean | Median | Standard Deviation |
|---|---|---|---|
| Accuracy | 0.9304 | 0.9397 | 0.0395 |
| Recall | 0.9738 | 0.9878 | 0.0381 |
| Precision | 0.9502 | 0.9540 | 0.0202 |
| F1 Score | 0.9684 | 0.9685 | 0.0055 |

Moreover, we study the system performance when utilizing dynamic deformation categorization based approach used in EarDynamic and static geometry based approach to generate templates for authentication. The static geometry-based approach generates the template when the user is under the static scenario (i.e, the user is not speaking). In this case, the generated template only represents the physiological characteristics of the ear canal. Using only static geometry of the ear canal provides a very good accuracy when the user is not speaking or moving the head. However, the ear canals of the user become dynamic when he/she is issuing voice commands or talking over the phone. Such dynamics will cause ear canal deformation and degrade authentication performance. Thus, the authentication system only utilizes the static information of the ear canal may not be sufficient under dynamic scenarios.

As shown in Fig. 16 under the first group of the bars "Static vs Static", enrolling and authenticating the user with static features in a static scenario result in the best performance, which indicates that static templates work fine in static scenarios. However, it is unrealistic to assume that the user's ear canals are always in static status. Indeed, the user's ear canals are dynamic while the user is speaking and moving his/her head. In such cases, when we use the static features to authenticate dynamic ear canals, the authentication performance plummets as shown in the third group of "Static VS Dynamic". To solve this problem, we introduced phoneme categories and build dynamic templates based on them, then use our dynamic ear canal template for the dynamic scenarios, and as the second group "Dynamic VS Dynamic" shows, authenticating with dynamic features achieve comparable results with first group. According to the results, we observe that although static templates work fine under a static scenario, only relying on the static template is not sufficient. Therefore, we believe that categorizing ear deformation by phonemes and building the dynamic ear canal profiles for the user could further enhance the performance and usability of our system.

Moreover, as shown in Fig. 17, during the authentication process, the correlation between extracted live user feature and dynamic deformation categorization based template is over 95% while the correlation between live user feature and the static geometry-based template is much lower. This shows our system can capture both the physiological and behavioral characteristics of the ear canal and defend against mimic attack.

### 5.3 Performance under Advanced Attack

Next, we evaluate the performance of our system under the advanced attack. To launch such an attack, the adversary leverages the leaked ear canal static geometry information of the victim and rebuilds the ear canal model. For this type of attacks, the attacker can only replicate the static geometry of the victim's ear canal at best and miss the dynamic deformation motion information when the user is speaking. Thus, to simulate the advanced attack, we ask the participant to wear the in-ear devices and replay their voice commands from other devices. The user will keep silent during the authentication process. As shown in Fig. 18, the blue curve is the correlation of the live user extracted feature with respect to the template and the red curve is the correlation of the advanced attack scenario extracted feature with respect to the template. We can observe that the correlation of the live user with respect to the template is consistently higher during the authentication process. This is because the advanced attack scenario can only capture the static geometry feature of the ear canal while missing the dynamic deformation motion information. Moreover, the false accept rate under the advanced attack is 5.3% and the false reject rate is 2.9%. This is because the dynamic deformation can only be captured through live user and can not be spoofed by the advanced attack. Such results show our system is effective in defending against advanced attacks.

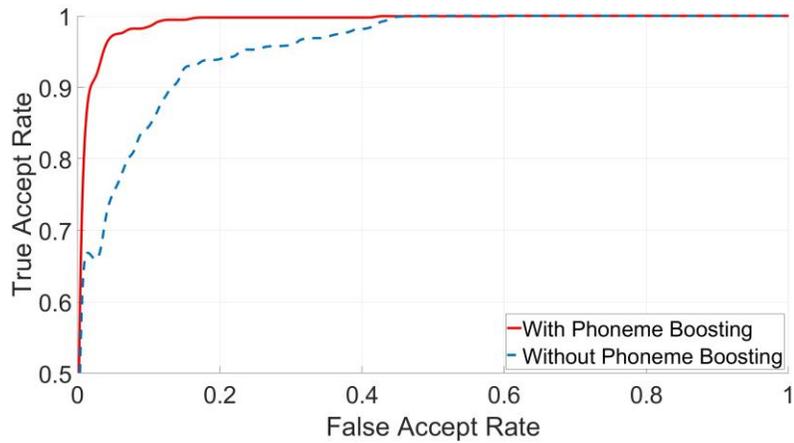

Fig. 14. ROC curves under without and with boosting.

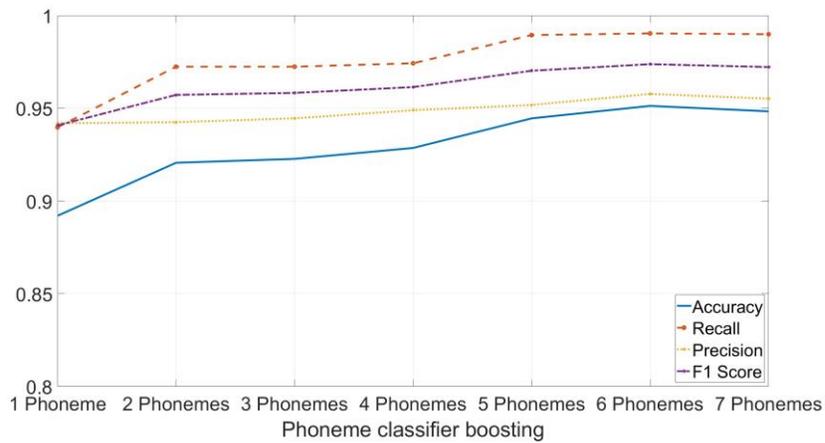

Fig. 15. Classifier boosting with multiple phoneme.

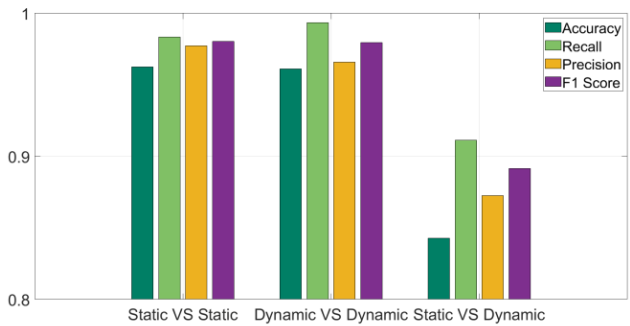

Fig. 16. Performance comparison of using static template and deformation-based template for user authentication.

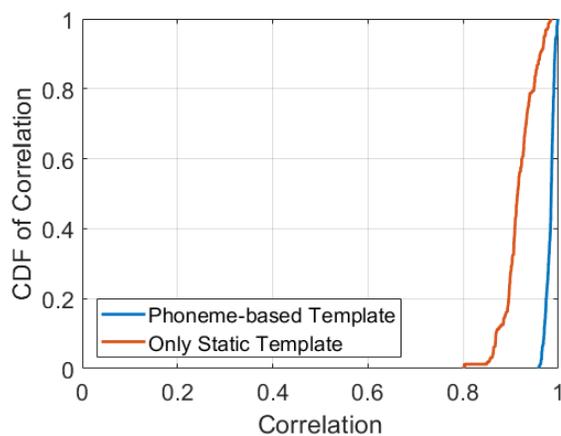

Fig. 17. Correlation comparison between deformation-based template and only using static template.

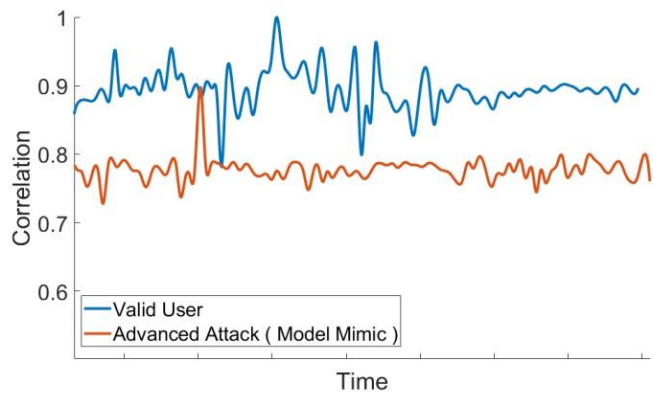

Fig. 18. Correlation of ear canal features compairson between valid user and model mimic attack.

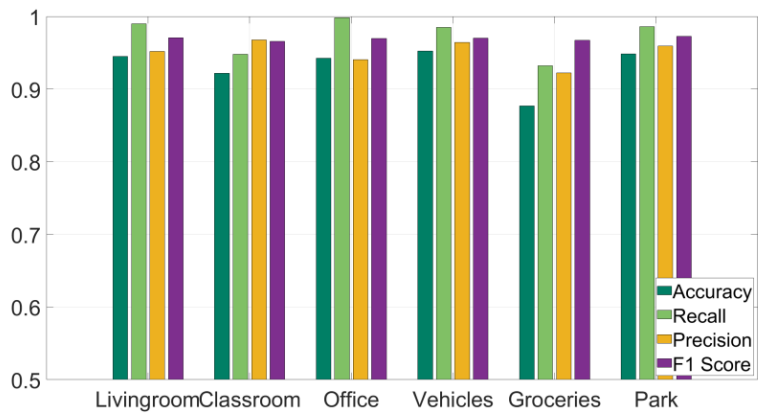

Fig. 19. Authentication performance in different environments with different background noise.

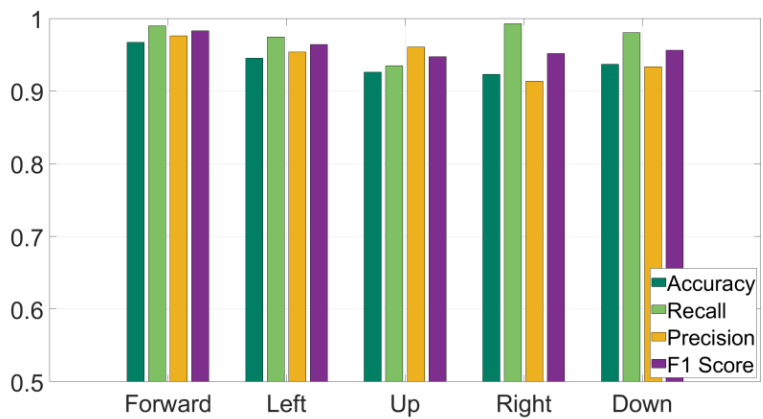

Fig. 20. Authentication performance with different head posture.

### 5.4 Impact of Different Environments

As people may wear in-ear devices at various locations, here we study our system's performance in different environments. We choose six typical environments: living room, classroom, office, vehicles, grocery store, and park. Fig. 19 shows that our system performance in the abovementioned environments. We observe that our system still works well even across the different environments. In particular, the accuracy of our system in six different environments are 94.50%, 92.17%, 94.25%, 95.25%, 87.70%, 94.83%, respectively. Moreover, we observe that all the environments besides the grocery store achieve comparable performance. Although the accuracy is slightly lower in the grocery store due to substantial background noises within the environment, our system still provides high authentication accuracy. Such observations show that our system is robust and can work well in different environments.

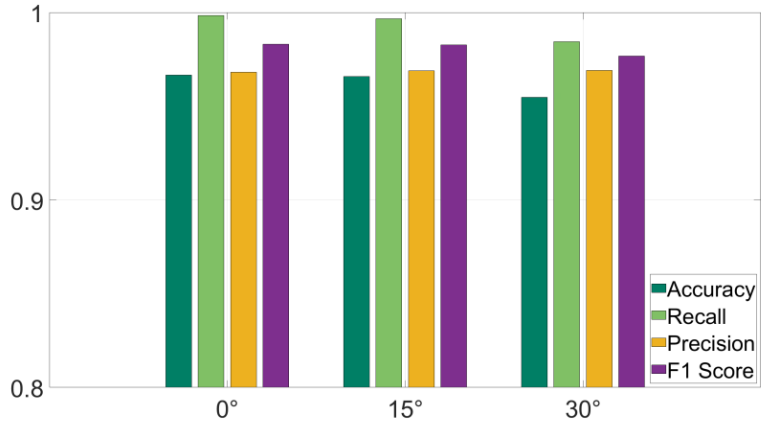

Fig. 21. Authentication performance with different wearing angles.

## 5.5 Impact of Head Posture

Difference head postures will also impact the ear deformation, thus we next study how these head postures impact our system. The head postures we consider is facing left, up, right, and down and using facing forward as our baseline. We use a motion sensor that is attached to the earbud to detect the head postures and decided to match with which template. As shown in Fig. 20, comparing with facing forward, the accuracy drop 2% to 4% for other head postures. Although compare with facing forward, a slight drop in the accuracy could be seen, our system still has good authentication accuracy under different head posture. This observation demonstrates that our system could work well when the users are speaking with different head postures.

## 5.6 Impact of Wearing Positions

People usually wear earbuds in their habitual position, it is possible the relative position of the earbud with respect to the ear could be slightly different from time to time. Next, we evaluate the wearing position of the in-ear devices' impact on our system performance. We experiment with three different wearing angles with respect to the normal wear position: 0°, 15°, and 30°. We collect additional 160 sentences for case of 15°, and 30° and 240 datapoints are used in total. The measured angle is facing toward the earlobe with respect to the original position and 0° is used to establish the baseline performance. The results are shown in Fig. 21. We observe that our system achieves similar performance across different wearing angles. In particular, the accuracy for 15° and 30° wearing angles are over 95%. This demonstrates our system is not sensitive to different wearing positions.

## 5.7 Impact of Time

To evaluate the robustness of our system over time, we ask some participants to use our system over various time periods after initial enrollment. The time periods we consider are 1 day, 10 days, 30 days, and 120 days. 140 additional sentences are sampled for 30 days and 120 days, 300 sentences are analyzed in total. As shown in Fig. 22, we can observe our system can maintain high performance over various time periods. In particular, our system achieves accuracy over 95% after 120 days. This is due to the consistency of the ear canal dynamic deformation and static geometry for adults. However, it is worth noticing that the static geometry of the ear canal will change gradually when the user is aging. Thus it is important to update the template if the user plans to use the system for an extended period of time (e.g., over one year).

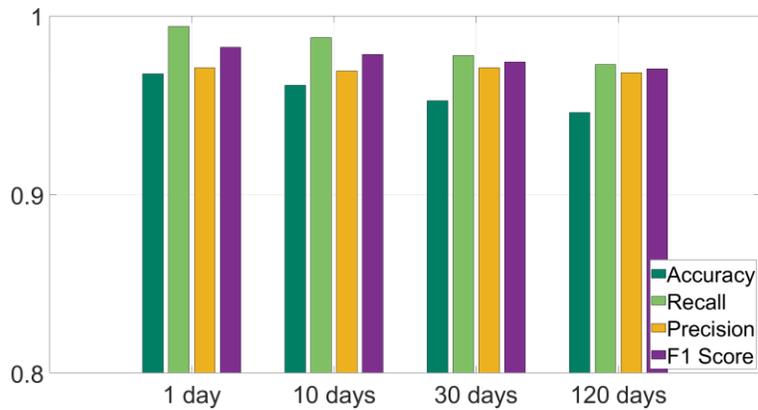

Fig. 22. Authentication performance over different time of period.

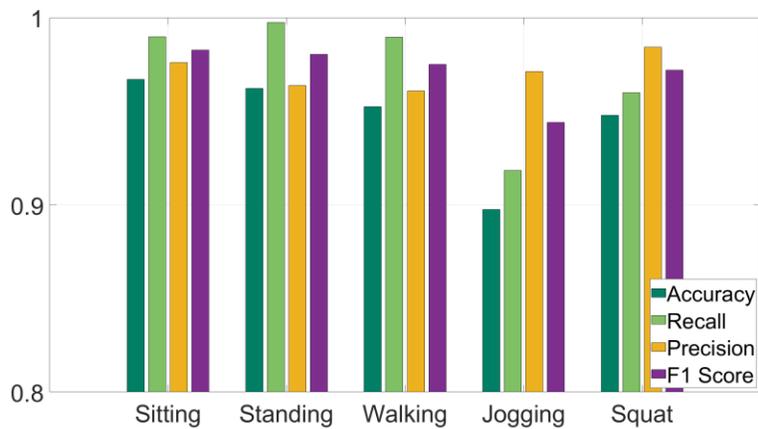

Fig. 23. Authentication performance with different body motions.

## 5.8 Impact of Body Motion

As we know different gestures could make an impact on the ear canal, here we study how different body motions will affect the performance of our system. The motions we chose are daily activities including sitting, standing, walking, jogging, and squat. For walking, jogging, and squat, we collected 360 sentences in addition. As shown in Fig. 23, our system can maintain high accuracy when the users are performing different body motions. Although the performance of our system drops during jogging, it can still achieve around 90% accuracy. It is possible the heavy breathing during jogging will likely change the way users speak. Therefore, the dynamic deformation motion when pronouncing the same phoneme under the jogging scenario is different and leads to an accuracy drop. Nevertheless, the results, in general, show that our system is robust to various body motions.

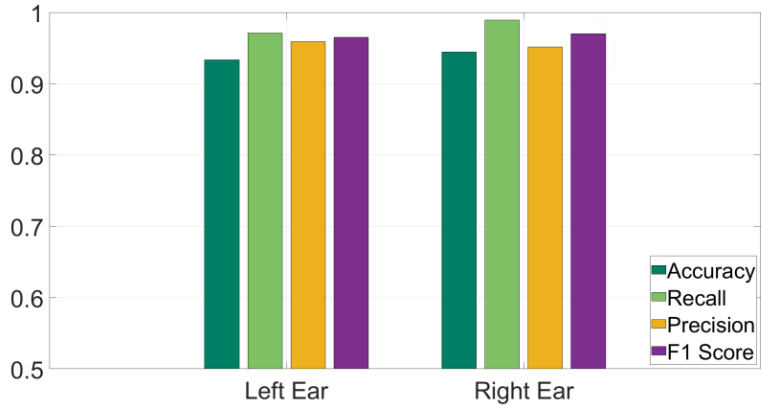

Fig. 24. Authentication performance based on left ear or right ear.

## 5.9 Study of Left and Right Ear

Next, we study the system performance over different ears. We ask the participants to use either left or right ear to enroll in the system and then utilize the corresponding ear for authentication. The results are shown in Fig. 24. Overall, our system is effective in authenticating the user using either ear. In particular, the accuracy and precision using the left ear are 93.31% and 95.58%. For the right ear, the accuracy is 94.45% and precision is 95.17%. In this work, we primarily utilize one ear (i.e., left ear) for experimental evaluation if not specified. It is worth mentioning when the user is wearing both earbuds, it provides us the opportunity to further enhance the system security by using the other ear as a secondary authentication factor.

## 5.10 User Study

We next perform a user study based on our dataset on different factors including gender, accent, and age.

**Gender.** We first study the impact of different gender. We recruit 24 participants including 12 females and 12 males. As shown in Fig. 25, the average accuracy, recall, precision, and F1 score for female are 92.24% 97.12% 94.27%, and 96.72%, respectively. And for male, they are 94.16% 97.75% 96.07%, and 97.00%, respectively. These results suggest that, in general, our system yields better performance on male users. We observe that this could be caused by the different anatomical composition and speaking habits, e.g., the male user usually has louder volume and bigger articulator gestures compared with the female users.

**Accent.** Both native English speakers and non-native English speakers are recruited for the accent experiments. In particular, there are 4 Native English speakers and 20 non-native English speakers from different countries. According to the Fig. 26, the average accuracy, recall, precision, and F1 score for native speakers are 93.48% 98.97% 94.18%, and 96.52%, whereas for non-native speakers, they are 92.95% 97.07% 95.19%, and 96.90% , respectively. Indeed, comparing with native English speakers, non-native speakers' pronunciations are less stable. For example, non-native speakers are prone to mispronounce phonemes. Such inconsistency might impact the authentication and lower our system performance.

**Age.** We next study how age could affect our system performance. Our participants are divided into four age categories: 20-25, 25-30, 30-35, and 35-40. Fig. 26 shows the average accuracy, recall, precision, and F1 score for these age groups. We observe that our system provides stable appropriate performance for all the age groups from 20 to 40. Specifically, the authentication accuracies for all age groups are above 93% and the variances are less than 1%.

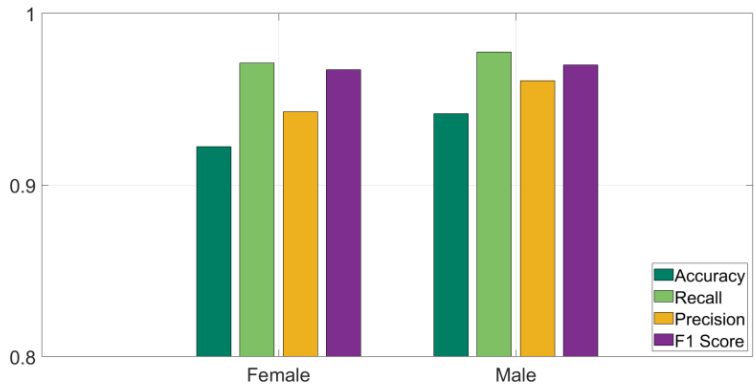

Fig. 25. User study on gender.

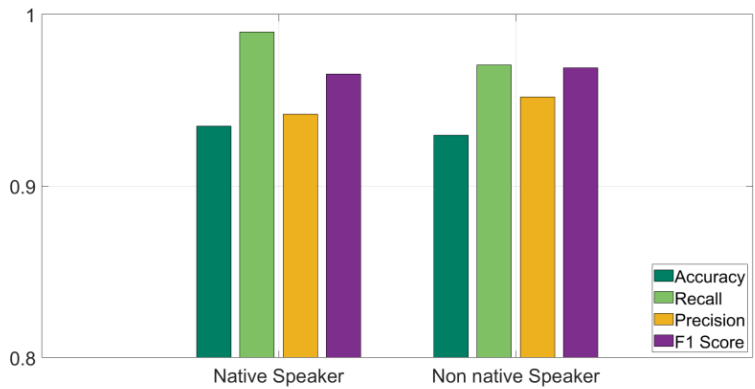

Fig. 26. User study on accent.

## 6 DISCUSSION

**Generalization to other languages.** Our system is currently built on the phonetics characteristics of the English language. Thus, it works for English or other languages that share similar phonetics features. Since the phonetics of most languages have the phonemes as their basic component, we could further analyze the phonetics of other languages and upgrade our model. For instance, the Japanese adopt a fifty sounds Syllabary called "Gojūon". All these sounds are all built on the five essential phonemes: a, i, u, e, o. Therefore, to apply our system to the Japanese language, we could generate a five categories based phoneme model with similar procedures aforementioned. Similarly, we believe with further phonetic analysis and phoneme categorization with respect to individual language, EarDynamic could be applied to other languages beyond English.

**Age and accents.** Also, since the phoneme categorization in our system is based on both articulatory phonetics and our experimental observations, the system might work best with the standard accent like Received Pronunciation (RP) and the General American (GA). Nevertheless, our user study on accents show that our system is robust against different foreign accents. Moreover, we could enlarge our datasets in the future to further improve our system's tolerance to variations like strong accents. We will include this in our future work.

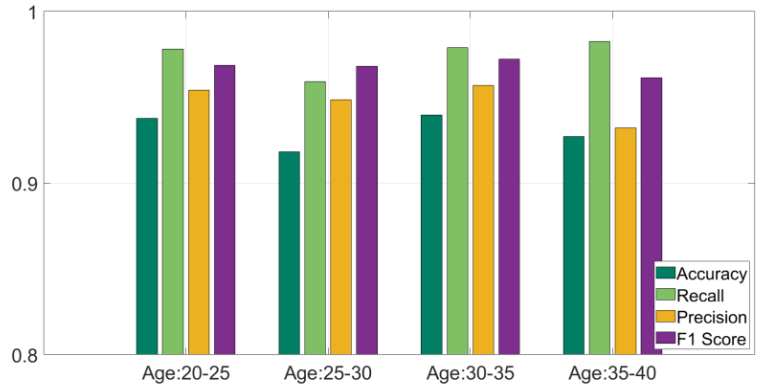

Fig. 27. User study on age.

**Ear infection and cold.** Besides, other factors like ear infection and cold could have an impact on the performance of our system. For instance, a severe ear infection in the ear canal (swimmer' ear) could make the ear canal swell and even affect the hearing ability of the patient [48], thus such swell in the ear canal could impact our system. The effect of cold could be the changes of the speaking voices due to the sore throat or clogged nasal. But it won't have many impacts on the articulator motions during the pronouncing process. Thus, we believe the cold will have a limited impact on our system. Furthermore, the static geometry of the ear canal will change slowly over the years. For instance, the canal could become more bent and narrowed when people grow older. Thus, it is beneficial to perform a long-term study across several years and sampling more data to periodically adjust the user profile.

**Facial expression.** Facial expression could also cause ear canal deformation [5]. Indeed, we may improve our system performance by including the ear canal deformations caused by facial expressions in our system design. However, it could be challenging to distinguish different types of facial expressions in practice. One possible scenario is to leverage the motion sensors to detect various facial expressions since each facial expression could have an unique impact on the motion sensors attached to the earbuds. We would like to include this as part of our future work.

**User Study.** Currently, due to the impact of COVID-19 pandemic, only a limited number of subjects participating in our experiments. A more comprehensive study over a larger user group could better evaluate our system. Moreover, the prototype we used in our experiment is hand made, an integrated hardware design will be better suited for large-scale deployment.

## 7 CONCLUSIONS

In this work, we propose EarDynamic, a continuous and passive user authentication system that leverages the ear canal deformation sensed by the in-ear wearable. Our study shows that the ear canal deformation due to articulation activities is unique for each individual and contains both the static geometry and dynamic motion of the ear canal when the user is speaking. We sense the ear canal deformation with an acoustic based approach that utilizes the microphone and speaker on the in-ear wearable. We also build a prototype of EarDynamic with off-the-shelf accessories by embedding an in-ward facing microphone inside an earbud. Extensive experiment results show that EarDynamic is highly accurate in authenticating users. Results also show that the system performs well under different noisy environments with various daily activities.